\documentclass[aps,prc,twocolumn,superscriptaddress,showpacs,eqsecnum]{revtex4}
\usepackage{graphicx}
\usepackage{dcolumn}
\usepackage{bm}

\newcommand{\rgn}{($\gamma$,n)}
\newcommand{\rng}{(n,$\gamma$)}
\newcommand{\rga}{($\gamma$,$\alpha$)}
\newcommand{\rgao}{($\gamma$,$\alpha_0$)}
\newcommand{\rag}{($\alpha$,$\gamma$)}
\newcommand{\rago}{($\alpha$,$\gamma_0$)}
\newcommand{\ran}{($\alpha$,n)}
\newcommand{\rgp}{($\gamma$,p)}
\newcommand{\rpg}{(p,$\gamma$)}

\newcommand{\spro}{$s$-process}
\newcommand{\rpro}{$r$-process}
\newcommand{\ppro}{$p$-process}
\newcommand{\pnuc}{$p$-nuclei}
\newcommand{\astph}{astrophysical}

\begin{document}

\title{
Photon-induced Nucleosynthesis:
Current Problems and Experimental Approaches
}

\author{P.\ Mohr}
\email{WidmaierMohr@compuserve.de}
\affiliation{
  Strahlentherapie, Diakoniekrankenhaus Schw\"abisch Hall, \\
  D--74523 Schw\"abisch Hall, Germany
}

\author{Zs.\ F\"ul\"op}
\affiliation{
  ATOMKI, P.O.\ Box 51, H-4001 Debrecen, Hungary
}
\author{H.\ Utsunomiya}
\affiliation{
  Department of Physics, Konan University, \\
  8-9-1 Okamoto, Higashinada, Kobe 658-8501, Japan
}
\date{\today}

\begin{abstract}
Photon-induced reactions play a key role in the nucleosynthesis of
rare neutron-deficient \pnuc . The paper focuses on \rga , \rgp , and
\rgn\ reactions which define the corresponding \ppro\ path. The
relation between stellar reaction rates and laboratory cross sections
is analyzed for photon-induced reactions and their inverse capture
reactions to evaluate various experimental approaches. An improved
version $S_C(E)$ of the astrophysical S-factor is suggested which is
based on the Coulomb wave functions. $S_C(E)$ avoids the apparent
energy dependence which is otherwise obtained for capture reactions on
heavy nuclei. It is found that a special type of synchrotron radiation
available at SPring-8 that mimics stellar blackbody radiation at
billions of Kelvin is a promising tool for future experiments.  By
using the blackbody synchrotron radiation, sufficient event rates for
\rga\ and \rgp\ reactions in the \ppro\ path can be expected. These
experiments will provide data to improve the nuclear parameters
involved in the statistical model and thus reduce the uncertainties
of nucleosynthesis calculations.
\end{abstract}



\maketitle

\section{Introduction}
\label{sec:intro}
Nucleosynthesis of heavy nuclei proceeds mainly via neutron capture
reactions and subsequent $\beta$-decays in the so-called \spro\ and
\rpro\ \cite{wallerstein}.
About 99\,\% of the heavy nuclei 
above iron are synthesized
in these processes. However, there are 35 nuclei on the
neutron-deficient side of the chart of nuclides, the so-called \pnuc , 
which cannot be made by
neutron capture. The dominant reactions for the synthesis of \pnuc\ 
are photodissociations induced by photons
of a hot stellar environment with temperatures around $1.8 \le T_9 \le
3.3$ (where $T_9$ is the temperature in $10^9$\,K) or thermal energies
155\,keV $\le kT \le$ 285\,keV. The required seed nuclei must have
been synthesized earlier in the {\it s--} and/or \rpro\
\cite{Ito61,Arn76,Woo78,Ray90,Pra90,How91,Lam92,Ray95,Rau02,Arn03,Fuj03,Hay04,Hay06,Uts06,Rapp06}.
The oxygen- and neon-rich layers of type II supernovae are good
candidates for the \astph\ site of the \ppro\ (note that ``{\it{p}}''
should be interpreted here as photodissociation, not as proton capture);
however, there is no firm conclusion on the \astph\ site yet
\cite{Arn03}.

The full reaction network for the nucleosynthesis calculations in the
\ppro\ contains by far more than 1000 nuclei and more than 10000
reactions, involving many unstable nuclei. It is impossible to measure
all reaction rates in the laboratory. Usually, the reaction rates are
calculated within the framework of the statistical model \cite{Rau00}.
It is the aim of this paper to discuss how the reliability of these
calculations can be tested using new experimental data which can be
obtained with synchrotron based photon sources at SPring-8
\cite{Tan02} or at other facilities like S-DALINAC at TU Darmstadt
\cite{TUD}, ELBE at Forschungszentrum Rossendorf \cite{Tei03}, or
HI$\gamma$S at TUNL, Duke University \cite{Wel03}.

The dominant reactions in the \ppro\ path are \rgn\ and \rga\
reactions leading to a path of the \ppro\ which is located about 10
mass units from stability on the neutron-deficient side around $A
\approx 200$ and close to stability around $A \approx 140$
\cite{Woo78,Arn03}. At lower masses around $A \approx 100$, the
importance of \rgp\ and \rpg\ reactions increases. Recently it has
been pointed out, that a fast expansion of high-entropy, proton-rich
matter may produce a significant amount of $^{92,94}$Mo and
$^{96,98}$Ru \cite{Jor04} which have been underestimated in most
previous calculations; but unfortunately no \astph\ site could be
assigned firmly to this scenario \cite{Jor04}.  In addition,
neutrino-induced nucleosynthesis has been suggested recently for the
synthesis of light \pnuc\ \cite{Pru06,Fro06}.

Whereas a lot of work has been done for \rgn\
reactions in recent 
years, there are still only few experimental
data for \rgp\ and especially for  \rga\ reactions at all, and only
very few data at \astph ly relevant energies. 

The paper is organized as follows. In Sect.\ \ref{sec:present} the
present status for \rga , \rgp , and \rgn\ reactions and indirect
approaches are discussed.  An improved treatment of the \astph\
S-factor is suggested which avoids an apparent energy dependence of
the conventional S-factor based on the Gamow factor for capture
reactions on heavy nuclei.  Sect.\ \ref{sec:inverse} presents a
detailed comparison between photodissociation reactions and capture
reactions in the laboratory and in hot stellar environments with a
focus on \rga\ and \rag\ reactions. In Sect.\ \ref{sec:spring8} as an
example the feasibility of \rga\ experiments at SPring-8 is analyzed,
and finally conclusions are given in Sect.~\ref{sec:conc}.

\section{Photodissociation for the Astrophysical p-process}
\label{sec:present}
In the following section we analyze the \astph ly relevant energy for
photon-induced reactions. The most effective energy $E_{\rm{eff}}$ and
the width $\Delta$ of the Gamow-like energy window are derived from
the thermal photon spectrum and from the energy
dependence of the cross sections of the respective photon-induced \rga
, \rgp , and \rgn\ reactions. This energy dependence can be estimated
from the relation between photodissociation reactions and their
inverse capture reactions. In addition, 
a brief overview of the available
experimental data in this energy region is given.

\subsection{\rga\ Reactions}
\label{subsec:basic}
Almost no experimental data exist for \rga\ reactions in the \astph ly
relevant energy window that is defined in the following way. The
photon density at temperature $T$ is given by the blackbody radiation:
\begin{equation}
n_\gamma(E,T) = 
  \left( \frac{1}{\pi} \right)^2 \,
  \left( \frac{1}{\hbar c} \right)^3 \,
  \frac{E^2}{\exp{(E/kT)} - 1}
\label{eq:planck}
\end{equation}
The stellar reaction rate of a \rga\ reaction is obtained by
\begin{equation}
\lambda^\ast(T) =
  \int_0^\infty 
  c \,\, n_\gamma(E,T) \,\, \sigma^\ast_{(\gamma,\alpha)}(E) \,\, dE
\label{eq:rate}
\end{equation}
with the cross section $\sigma^\ast$ under stellar conditions, i.e.,
including thermal excitations of the target nucleus (see also Eqs.~(1)
and (2) in \cite{Rau00}). For completeness,
we define here also the reaction rate in the laboratory
$\lambda^{\rm{lab}}$ with the target in the ground state (see also
Sect.~\ref{sec:spring8})
\begin{equation}
\lambda^{\rm{lab}}(T) =
  \int_0^\infty 
  c \,\, n_\gamma(E,T) \,\, \sigma^{\rm{lab}}_{(\gamma,\alpha)}(E) \,\, dE
\label{eq:ratelab}
\end{equation}
where $\sigma^{\rm{lab}}$ is the ground state cross section. The rate
$\lambda^{\rm{lab}}$ can be measured in the laboratory using photon
spectra with a (quasi-) thermal energy distribution which can be
obtained by a superposition of bremsstrahlung spectra in a narrow
energy window \cite{MohrPLB,Uts06} or by high-energy synchrotron
radiation of GeV electron beams in a much broader energy window
\cite{Uts05_NIMA}. For simplicity of the following notation, we use the
symbol $\lambda$ also for the reaction rate of the 
inverse \rag\ capture
reaction: 
\begin{equation}
\lambda_{(\alpha,\gamma)} := \langle \sigma \, v
\rangle_{(\alpha,\gamma)}
\label{eq:def}
\end{equation}

The energy dependence of the \rga\ cross section can be derived from
the inverse \rag\ cross section. The \rag\ capture reaction shows
a roughly exponential energy dependence at low energies because of the
tunneling probability through the Coulomb barrier. The nuclear part of
the energy dependence can roughly be seen in the \astph\ S-factor
defined by 
\begin{equation}
\sigma(E) = 
	\frac{1}{E} \, \exp{(-2 \pi \eta)} \, S(E)
\label{eq:sfact}
\end{equation}
with the Sommerfeld parameter $\eta$. The S-factor has a much weaker
energy dependence for capture reactions than the cross section. 


In the definition of the astrophysical S-factor only incoming
$s$-waves are taken into account. However, the influence of the
centrifugal barrier for higher partial waves with angular momenta $l >
0$ is small for reactions between heavy nuclei at energies around $5 -
10$\,MeV.  Thus, the energy dependence of the cross section for
capture reactions, in particular, from 
the incoming $p$-wave, is close to $s$-wave capture.

\begin{figure}[hbt]
\includegraphics[ bb = 125 90 535 660, width = 8.0cm, clip]{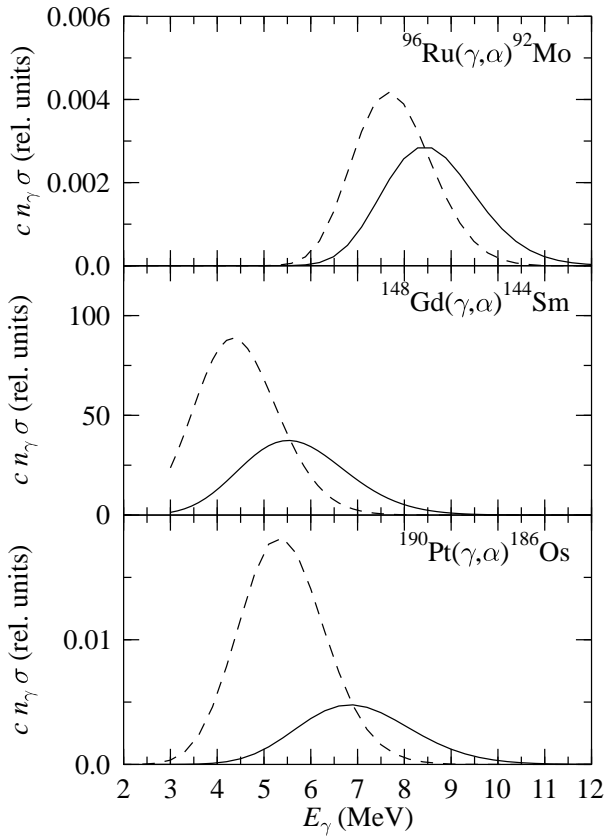}
\caption{
\label{fig:gamow_svar}
Gamow window of selected \rga\ reactions
around masses $A \approx 100$, 150, and 200: 
the integrand of
Eq.~(\ref{eq:ratelab}) is shown ($i$) using a constant S-factor of the
inverse \rago\ capture reaction (full lines) and ($ii$) using the
theoretical \rga\ cross
sections of \cite{Rau04} (dashed lines) which correspond to the
energy-dependent S-factors of Fig.~\ref{fig:sfaktvar}. 
The energy dependence of the S-factor leads to a
shift of the Gamow window by typically a few hundred keV up to about
1500\,keV. Note that the absolute scale of the data is arbitrary,
though it reflects the S-factor dependence of the reaction.
}
\end{figure}

In a very first approximation one
may assume a constant S-factor of the inverse \rago\ capture reaction
to the ground state.
One finds a strict correspondence between the \rago\ and \rgao\ cross
sections where the $p$-wave $\alpha$ capture on even-even nuclei
populates 1$^{-}$ states in the compound nuclei that decay directly to
the ground state (0$^{+}$) by emitting E1 photons. 
By applying the reciprocity theorem which relates the \rgao\ and \rago\ 
cross sections in the laboratory (see Sect.~\ref{sec:inverse}), 
one finds that the integrand of
Eqs.~(\ref{eq:rate}) and (\ref{eq:ratelab}) shows a maximum at
\begin{equation}
E_{\rm{eff}}^{(\gamma,\alpha)} \approx 0.122 \, \times \, \left(Z_P^2
\, Z_T^2 \, A_{\rm{red}} \, T_9^2\right)^{1/3}\,{\rm{MeV}} + Q_\alpha
= E_{\rm{eff}}^{(\alpha,\gamma)} + Q_\alpha
\label{eq:gamow_alpha}
\end{equation}
where $Z_P$, $Z_T$, and $A_{\rm{red}}$ are the charge numbers of
projectile and target and the reduced mass number of the inverse \rag\
reaction, and $Q_\alpha$ is the $Q$-value of the \rag\ capture
reaction which corresponds to the binding energy of the $\alpha$
particle in the compound
nucleus. Note that $Q_\alpha < 0$ for many $p$ nuclei. The integrand
of Eq.~(\ref{eq:ratelab}) and its maximum are
shown in Fig.~\ref{fig:gamow_svar}. The width of this maximum is
approximately given by 
\begin{equation}
\Delta^{(\gamma,\alpha)} \approx 0.237 \, \times \, \left(Z_P^2 \,
Z_T^2 \, A_{\rm{red}} \, T_9^5\right)^{1/6}\,{\rm{MeV}}
\label{eq:width}
\end{equation}

The Gamow-like window of the \rga\ reactions is approximately
identical to the Gamow-window of the \rag\ capture reaction, but
shifted by the binding energy $Q_\alpha$
\cite{MohrNPDC,MohrTours}. The reason for this similarity is that for
both \rga\ and \rag\ reactions the most effective energy is governed by
the tunneling through the Coulomb barrier in the entrance channel 
in \rag\ and in the exit channel in \rga . Consequently, the position
and width of the Gamow-like window for \rga\ reactions are
temperature-dependent, and the width is much broader for \rga\
reactions compared to \rgn\ reactions. As an example, the position and
width for the Gamow-like window of the reactions $^{148}$Gd\rgn
$^{147}$Gd ($S_{\rm{n}} = 8984$\,keV) and $^{148}$Gd\rga $^{144}$Sm
($Q_\alpha = -3271$\,keV) are listed in Table \ref{tab:gamow}. This
example has been chosen because the nucleus $^{148}$Gd is a branching
point in the \astph\ \ppro\ which may be used as \ppro\ chronometer
\cite{Woo78,Arn03,Som98,Rau95,Rau06}. 
\begin{table*}[htb]
\caption{\label{tab:gamow} Most effective energy $E_{\rm{eff}}$ and
width $\Delta$ of the Gamow-like window of the reactions
$^{148}$Gd\rgn $^{147}$Gd ($S_{\rm{n}} = 8984$\,keV) and
$^{148}$Gd\rga $^{144}$Sm ($Q_\alpha = -3271$\,keV), derived under the
assumption of a constant S-factor of the inverse \rago\ capture
reaction. Note that the 
definitions for the widths are not fully consistent: for \rgn\
reactions $\Delta_{\rm{FWHM}}$ is given \cite{MohrNIC2000}, whereas
for \rga\ reactions the usual definition for $\Delta_{1/e}$ is used
\cite{Rolfs}. Because of the asymmetry of the integrand in
Eq.~(\ref{eq:rate}) there is no precise relation between
$\Delta_{\rm{FWHM}}$ and $\Delta_{1/e}$, but
$\Delta_{\rm{FWHM}}^{(\gamma,\alpha)} \approx 0.833 \,
\Delta_{1/e}^{(\gamma,\alpha)}$ is a rough approximation for the less
asymmetric \rga\ case. Additionally, the effective energy
$E_{\rm{eff}}^{(\alpha,\gamma)}$ for the \rag\ capture reaction is
given. For the widths: $\Delta_{1/e}^{(\alpha,\gamma)} \approx
\Delta_{1/e}^{(\gamma,\alpha)}$.
} 
\renewcommand{\tabcolsep}{0.75pc} 
\begin{tabular}{@{}cccccccc}
\hline
$T_9$	& $kT$\,(keV)
& $E_{\rm{eff}}^{(\gamma,{\rm{n}})}$\,(MeV)	
& $\Delta_{\rm{FWHM}}^{(\gamma,{\rm{n}})}$\,(MeV) 
& $E_{\rm{eff}}^{(\gamma,\alpha)}$\,(MeV)	
& $\Delta_{1/e}^{(\gamma,\alpha)}$\,(MeV) 
& $E_{\rm{eff}}^{(\alpha,\gamma)}$ \\
\hline
2.0	& 172	& 9.07	& 0.32 & 4.30 & 2.64 & 7.57 \\
2.5	& 215	& 9.09	& 0.41 & 5.52 & 3.18 & 8.79 \\
3.0	& 259	& 9.11	& 0.50 & 6.66 & 3.70 & 9.93 \\
\hline
\end{tabular}
\end{table*}

The energy dependence of the astrophysical S-factor at low energies is
not well determined. It 
cannot be taken directly from experimental data. In most cases the
experimental S-factor decreases slightly with increasing energy
\cite{Som98,Gyu06,Oe02,Oe07,Rapp02,Ful96,Hah65,Dal68,Ver67}. However, the
experimental data usually do not cover the full Gamow window but have
been measured at somewhat higher energies. There is urgent need for
improved experimental \rag\ data at low energies. It has to be pointed
out that the experimental determination of \rag\ cross sections at
\astph ly relevant energies is very difficult. The cross sections are
very small around the effective energy $E_{\rm{eff}}^{(\alpha,\gamma)}$ 
because of the rapidly decreasing tunneling probability below the
Coulomb barrier. Nevertheless, a number of
reactions has been measured in the last years at energies close or
slightly above $E_{\rm{eff}}^{(\alpha,\gamma)}$ including the
following \pnuc~: $^{144}$Sm\rag
$^{148}$Gd \cite{Som98}, $^{112}$Sn\rag $^{116}$Te \cite{Oe02,Oe07},
$^{106}$Cd\rag $^{110}$Sn \cite{Gyu06}, $^{96}$Ru\rag $^{100}$Pd
\cite{Rapp02}. Further data for heavy nuclei exist for $^{70}$Ge\rag
$^{74}$Se \cite{Ful96}, $^{106}$Cd\rag $^{110}$Sn \cite{Hah65}, and
$^{139}$La\rag $^{143}$Pr \cite{Dal68,Ver67}, and extensive studies
are in progress \cite{har05}. 

It has been noticed especially for the case of the
$^{144}$Sm\rag $^{148}$Gd reaction that the theoretical prediction of
the capture cross section slightly below 10\,MeV (which is the \astph
ly relevant energy, see Table \ref{tab:gamow}) varied by about two
orders of magnitude, mainly depending on the $\alpha$-nucleus
potential \cite{Som98}.
Consequently, $\alpha$-nucleus potentials for heavy \pnuc\ were
studied using elastic scattering at energies around the Coulomb
barrier. Data for $^{144}$Sm \cite{Mohr97}, $^{92}$Mo \cite{Ful01},
and $^{112}$Sn \cite{Gal05} are available, and a reasonable
description of the scattering data and a variety of further
experimental data including reaction cross sections and
$\alpha$-decay half-lives have
been obtained using systematic folding 
potentials \cite{Atz96,Mohr00,Dem03,Avr06,Mohr07}. 
However, the low-energy behavior
of the potentials is not well-defined; this still leads to
considerable uncertainties for the prediction of cross sections at
\astph ly relevant energies. Further experimental data are
strongly required, as e.g.\ pointed out in \cite{Dem03}.
Another possibility for studying the $\alpha$-nucleus potential is using
the sensitivity of (n,$\alpha$) reactions, 
for which details can be found in \cite{Gle00,Koe04}.

There are also attempts for a comprehensive investigation of
alpha-induced reactions on the same nucleus. For example, for the
$^{106}$Cd nucleus ($\alpha$,$\gamma$), ($\alpha$,p) and ($\alpha$,n)
reactions have been measured and compared to statistical model
calculations with different input parameter sets \cite{Gyu06}. It is
found that there is no parameter set describing all three channels
simultaneously. This reflects the limited knowledge on the optical
potentials in the mass and energy region relevant to the astrophysical
\ppro .  Similar study is in progress for the $^{112}$Sn isotope
\cite{Ozkan06,Oe07}.

As already pointed out above,
theoretical predictions of \rag\ capture cross sections show a
dramatic variation in their energy dependence (see e.g.\ Fig.~10 of
\cite{Gal05} for $^{112}$Sn\rag $^{116}$Te and Fig.~10 of \cite{Dem03}
for $^{144}$Sm\rag $^{148}$Gd). To estimate the influence of an
energy-dependent S-factor on the position of the Gamow window of \rga\
photodissociation reactions, the following procedure 
can be applied. In a first step the statistical model \rga\ cross
sections compiled in \cite{Rau04} have been converted to the corresponding
laboratory \rago\ capture cross section to the ground state using
time-reversal symmetry.  This is a good approximation because of the
dominating ground state contribution in the \rga\ reaction (see
Sect.~\ref{sec:inverse}). It is noted that since \rga\ reactions are 
dominantly induced by E1 photons, the \rago\ reaction of time-reversal
symmetry is characterized by $p$-wave $\alpha$ capture on even-even nuclei.  
However, as far as the energy dependence 
is concerned, one expects essentially no difference between $s$-wave and 
$p$-wave capture because the effect of the centrifugal potential is 
very small in the energy region of interest.  

\begin{figure}[hbt]
\includegraphics[ bb = 145 60 490 255, width = 8.0cm, clip]{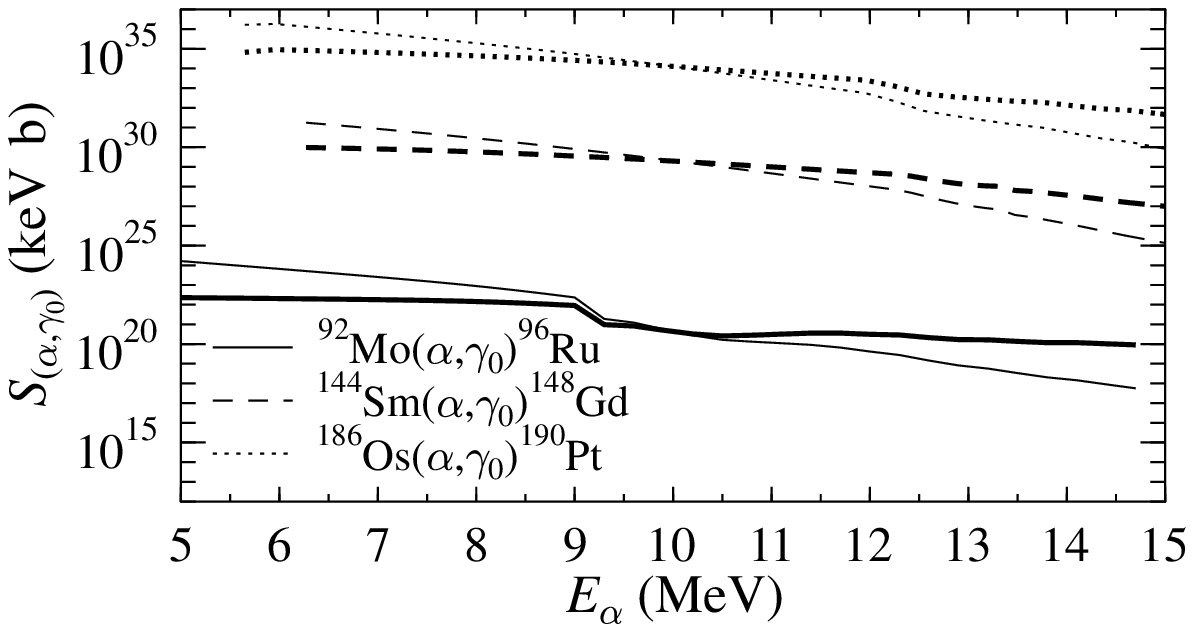}
\caption{
\label{fig:sfaktvar}
S-factor of the $^{92}$Mo\rago $^{96}$Ru, $^{144}$Sm\rago $^{148}$Gd, 
and $^{186}$Os\rago $^{190}$Pt calculated from the statistical model 
\rga\ photodissociation cross sections of \cite{Rau04}. 
The thin lines correspond to the
standard S-factor in Eq.~(\ref{eq:sfact}), the thick lines use the
new improved S-factor $S_C$ in Eq.~(\ref{eq:s_impr}). The new S-factor
$S_C$ shows a much weaker energy dependence than the standard S-factor
$S(E)$. The $S_C$ curves are shown in relative units and are normalized to
$S(E)$ at $E = 10$\,MeV. The chosen radius is $r_0 = 1.3$\,fm.
For $T_9 = 2.5$ the Gamow windows in Eq.~(\ref{eq:gamow_alpha}) are
located at $E_{\rm{eff}}^{(\alpha,\gamma)} = 6.75$\,MeV, 8.79\,MeV,
and 10.09\,MeV for the above capture reactions, respectively. 
The kinks in $S(E)$ and $S_C(E)$ at 9\,MeV for
the $^{92}$Mo\rago $^{96}$Ru reaction and at 12\,MeV for the
$^{144}$Sm\rago $^{148}$Gd and $^{186}$Os\rago $^{190}$Pt reactions
indicate the opening of the \ran\ channel; here the competition between
neutron emission and $\gamma$-ray emission reduces the \rago\ cross
section. 
}
\end{figure}

Three examples have been chosen for the analysis of the energy
dependence of the astrophysical S-factor $S(E)$ around $A
\approx 100$, $150$, and $200$: $^{92}$Mo\rag $^{96}$Ru,
$^{144}$Sm\rag $^{148}$Gd, and $^{186}$Os\rag $^{190}$Pt and their
inverse \rga\ photodissociation reactions. The \rago\ cross
sections which are derived from the theoretical \rga\ cross sections
in \cite{Rau04} have been translated to \astph\ S-factors which are shown in
Fig.~\ref{fig:sfaktvar}. The \astph\ S-factor decreases with energy;
the slope is about one order of magnitude per 2\,MeV within 
the Gamow window. 

The consequences of this energy dependence for the Gamow window of the
\rga\ cross section are shown in Fig.~\ref{fig:gamow_svar}. For the
three selected examples the integrand of Eq.~(\ref{eq:ratelab}) is
shown using a constant S-factor (full lines) and using the model
\rga\ cross sections of \cite{Rau04} (dashed lines) which correspond
to the energy-dependent S-factors of Fig.~\ref{fig:sfaktvar}. The
temperature for all figures is $T_9 = 2.5$ which is a typical
\ppro\ temperature. Using the theoretical energy dependence of the
\rga\ cross section \cite{Rau04},
one finds a shift of the Gamow window to lower energies
between a few hundred keV and up to about 1500\,keV for the
$^{190}$Pt\rga $^{186}$Os case. Although there is a significant change
in the energy dependence of the cross section, the resulting shift of
the energy of the Gamow window remains limited because the energy
dependence of the thermal photon density is by far dominating; e.g.,
the exponential factor $\exp{(-E/kT)}$ changes by about 10 orders of
magnitude between 5\,MeV and 10\,MeV whereas the variation of the
cross section is about 2 orders of magnitude when one replaces the
simple assumption of a constant S-factor with a realistic energy
dependence (see Fig.~\ref{fig:sfaktvar}).

\subsection{An improved treatment of the astrophysical S-factor}
\label{sec:sfact}
The Gamow factor $\exp{(-2 \pi \eta)}$ in the conventional definition of 
the astrophysical S-factor in Eq.~(\ref{eq:sfact}) presumably
approximates the Coulomb-barrier  
tunneling probability for $s$-wave particles in the absence of the
centrifugal potential. This probability, $P$, is expressed by 
$P = \mid \psi(R_N) \mid^2/\mid \psi(R_c) \mid^2$, where the
denominator is the  
probability of finding the particle at the classical
turning point $R_c$ and the numerator is that at the nuclear 
radius $R_N$. For the one-dimensional Schr\"odinger equation 
for the Coulomb potential $V_C$ one finds 

\begin{equation}
P = \exp{\Bigl[ -\frac{2}{\hbar} \, \int_{R_N}^{R_c} 
\sqrt{2\mu \, \bigl(\,V_C(r) - E\bigr)}\, dr\Bigr]}.
\label{eq:t_wkb}
\end{equation}
The WKB solution of the three-dimensional Schr\"odinger equation gives 
the same equation with an extra proportional coefficient $\sqrt{(B_C-E)/E}$
with the Coulomb barrier height $B_C$ \cite{Clayton}. 

At low energies $E \ll B_C$ 
(typically by a factor of the order of 100) or, equivalently, 
when the classical turning point is much larger than the nuclear
radius $R_c \gg R_N$, 
$P$ can be approximated by the Gamow factor $\exp{(-2 \pi \eta)}$. In
particular, Eq.~(\ref{eq:t_wkb}) becomes $P$ = $\exp{(-2 \pi \eta)}$
for (the unrealistic case) $R_N = 0$ (point-like nuclei).
In contrast, in the $\alpha$ capture reactions of current interest,
the effective energy is no longer much smaller 
than the Coulomb barrier height. For example, as seen in Table
\ref{tab:gamow}, the effective energy $E_{\rm{eff}}^{(\alpha,\gamma)}$ at
$T_9 = 2 -3$ is $7 - 10$\,MeV for $^{144}$Sm, while the Coulomb 
barrier height is around 20\,MeV with the radius parameter $r_0 = 1.3$\,fm in 
$R = r_0 (A_1^{1/3} + A_2^{1/3}$). 

\begin{figure}[hbt]
\includegraphics[ bb = 145 95 530 485, width = 8.0cm, clip]{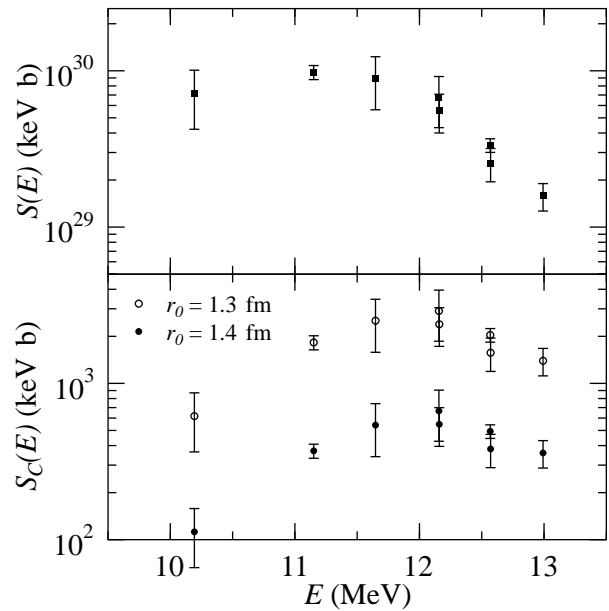}
\caption{
\label{fig:smagexp}
Experimental S-factor of the capture reaction $^{144}$Sm\rag
$^{148}$Gd \cite{Som98}, analyzed using the standard S-factor from
Eq.~(\ref{eq:sfact}) (upper part) and the new S-factor $S_C$
from Eq.~(\ref{eq:s_impr}) (lower part). Whereas the standard S-factor
decreases with energy by almost one order of magnitude between
11.5\,MeV and 13\,MeV, the new S-factor $S_C$ is constant within
a factor of two (except the lowest data point with its larger
experimental uncertainty).  To show the $r_0$ dependence of $S_C$,
two cases for $r_0 = 1.3$\,fm (open circles) and $1.4$\,fm (solid circles) 
are shown. Note that both diagrams cover a range of a factor of 50 for
$S(E)$ and $S_C(E)$ for better comparison.
}
\end{figure}

At large distances ($r \geq R_N$), 
where the nuclear potential is absent, the solution of 
the radial Schr\"odinger equation is given by a linear combination of the 
regular and irregular Coulomb functions, F$_{\ell}(r)$ and G$_{\ell}(r)$.  
(The assumption of the absence of the nuclear potential
holds for a potential well with a radius $R_N$, but obviously not for the 
Wood-Saxon potential or other realistic potentials.)
Thus, as it is well known, a better approximation of the $s$-wave tunneling 
probability is given with the radial wave function $u(r)$ \cite{Clayton} by
\begin{equation}
P = \frac{\mid u(\infty) \mid^2}{\mid u(R_N) \mid^2} = 
\frac{1}{F^2_0(E,R_N) + G^2_0(E,R_N)}. 
\end{equation}
Accordingly, a better astrophysical S-factor $S_C(E)$ is defined for
the present case:
\begin{equation}
\sigma(E) = 
	\frac{1}{E} \, \frac{1}{F^2_0(E,R_N) + G^2_0(E,R_N)} \, S_C(E)
\label{eq:s_impr}
\end{equation}
A minor disadvantage of the new $S_C$ is in the choice of the radius
parameter $r_0$ though the $r_0$ dependence of $S_C$ is moderate (see
Fig.~\ref{fig:smagexp}). We have chosen here $r_0 = 1.3$\, fm.

First we apply the new S-factor $S_C$ to the energy
dependence of the \rago\ cross section which is derived from the
calculation of \cite{Rau04}. The new S-factor is shown by the thick lines in 
Fig.~\ref{fig:sfaktvar} in comparison with the standard S-factor (thin lines).
One finds a much weaker energy dependence of $S_C(E)$ compared to $S(E)$.

Additionally we apply the definition of the new S-factor $S_C$ to
the experimental data of the $^{144}$Sm\rag $^{148}$Gd reaction
\cite{Som98} in Fig.~\ref{fig:smagexp}. Whereas the standard S-factor
shows a noticeable energy dependence (upper part), the experimental
data are almost constant when presented using the new $S_C(E)$. The influence
of the chosen radius parameter $r_0$ on the energy dependence of the
improved S-factor $S_C(E)$ is small.

The energy dependence of the \rag\ cross section derived from a constant
$S_C(E)$ is almost identical to that of the capture cross section 
converted from the statistical-model \rga\ cross section of
\cite{Rau04}. Or, in turn, one finds an almost energy-independent
S-factor $S_C(E)$ for the \rag\ cross section derived from
\cite{Rau04}. As can be seen from Fig.~\ref{fig:sfaktvar}, $S_C(E)$
varies typically by less than one order of magnitude within an energy
range of about 5\,MeV. (Note that the kinks in $S_C(E)$ at 9\,MeV for
the $^{92}$Mo\rago $^{96}$Ru reaction and at 12\,MeV for the
$^{144}$Sm\rago $^{148}$Gd and $^{186}$Os\rago $^{190}$Pt reactions
indicate the opening of the \ran\ channel.)
As a consequence of the similar energy dependencies, a similar shift
of the Gamow window to lower energies is found ($i$) from a model
prediction of the \rga\ cross section and time-reversal symmetry, and
($ii$) from an improved and almost energy-independent S-factor $S_C$
as defined in Eq.~(\ref{eq:s_impr}).

Eq.~(\ref{eq:gamow_alpha}) can thus provide only a rough estimate for
the position of the Gamow window for reactions between heavy
nuclei. Taking a more realistic energy dependence of the capture cross
section, e.g.\ from the improved and almost energy-independent
$S_C(E)$ or from the calculations of \cite{Rau04}, leads to a shift of
the Gamow window to lower energies by several hundred keV.

For completeness it has to be pointed out that the shift of the Gamow
window to lower energies applies not only to the \rga\
photodissociation, but also to the \rag\ capture reaction. In such
cases it is important to calculate \astph\ reaction rates $\langle
\sigma \, v \rangle$ by numerical integration of the energy-dependent
cross section instead to use simple approximations which are based on
the S-factor at the most effective energy $S(E_{\rm{eff}})$ and listed
in textbooks (e.g.~\cite{Rolfs}). The
slight shift of the Gamow window to lower energies does not affect the
comparison between \rag\ and \rga\ reactions in
Sect.~\ref{sec:inverse}.

Summarizing the above,
the observed noticeable energy dependence of the
astrophysical S-factor for \rag\ reactions discussed in the preceding 
subsection is a consequence of the definition of the S-factor with the
Gamow factor $\exp{(-2 \pi \eta)}$. The Gamow factor is valid only for $E
\ll B_C$ ($R_N \ll R_c$), 
which is not the case for \rag\ reactions of current interest.  
The new and improved S-factor $S_C$, which is based on the Coulomb
wave functions, is recommended to be used here. 
Its energy dependence is much weaker compared to the
conventional S-factor. Extrapolations of experimental data to lower
energies should be preferentially performed using the new S-factor
$S_C$ to avoid systematic errors.

\subsection{\rgp\ Reactions}
\label{subsec:rgp}
Most of the above arguments for \rga\ reactions in
Sect.~\ref{subsec:basic} hold for \rgp\ reactions, too. Recently, much
effort has been spent on \rpg\ capture reactions for nuclei with
masses around $A \approx 100$
\cite{gyu03,gyu01,ga03,ts04,la87,ch99,ha01,sa97,bo98,Gyu07a}, and the overall
agreement 
between these experimental data and theoretical predictions is
good. In addition, the theoretical estimates do not depend as strongly
on input parameter sets as in the case of \rag\ and \rga\ reactions.
The available datasets are compiled in a new on-line database 
\cite{kadonis}.

Experimental data for \rgp\ reactions, however, are rare \cite{Wag05}
and practically not available at \astph ly relevant energies. As will
be discussed in Sect.~\ref{subsec:rgp2}, better experimental data can 
reduce the uncertainties in the calculation of \astph\ \rgp\ reaction
rates.

\subsection{\rgn\ Reactions}
\label{subsec:rgn}
Recently, much effort has been spent on the analysis of \rgn\ reaction
rates. The \astph ly relevant energy window has been discussed first by
\cite{MohrPLB}; it is located close above the threshold of the \rgn\
reaction and has a typical width of less than 1\,MeV. 
The Gamow-like window is located at
$E_{\rm{eff}}^{(\gamma,{\rm{n}})} \approx S_{\rm{n}} + kT/2$, i.e.,
its position changes by less than 100\,keV between $T_9 = 2$ and $T_9
= 3$. The width 
increases slightly with
temperature from about 300\,keV at $T_9 = 2$ to about 500\,keV at $T_9
=3$ \cite{MohrNIC2000}.

Experimental data have been measured for a number of nuclei using the
photoactivation technique and a quasi-thermal photon spectrum that can
be obtained from the superposition of several bremsstrahlung spectra
\cite{MohrPLB}; the results are summarized in
\cite{Uts06,Sonn04}. Additionally, \rgn\ cross sections have been
measured using quasi-monochromatic photons from Laser Compton
scattering (LCS) \cite{Uts03,Shi05,Gok06}. It has been shown that the
experimentally obtained cross sections and reaction rates are in good
agreement with theoretical predictions. Typical deviations are much
smaller than a factor of two for the reactions investigated
\cite{Uts06,Sonn04}. 

The measurement of cross sections via Coulomb dissociation
\cite{baur} is a further 
method for the determination of photon-induced cross
sections. Here the strong virtual photon field of a heavy target
nucleus is used for the electromagnetic breakup of the projectile. 
The main advantage of
this method is the much larger cross section (compared to the direct
photon-induced reaction) which allows to study the properties of
unstable nuclei using radioactive ion beams \cite{Son06}.
However, one must be aware of such complications in 
the interpretation of data as contributions of nuclear breakup, 
a mixture of virtual photons with different multipolarity, 
and the final-state interaction known as post-Coulomb acceleration.

Further experimental data for \rgn\ reactions are
required for the comparison between theory and experiment in a broader
mass range and for a reliable extrapolation to unstable nuclei.

\section{Photodissociation and radiative capture in the lab and under
  stellar conditions}
\label{sec:inverse}
\subsection{\rag\ and \rga\ Reactions}
\label{subsec:rag}
For simplicity, the following
discussion is restricted to even-even nuclei, the most
interesting cases for \ppro\ nucleosynthesis \cite{Woo78}. The
discussion is illustrated using the example of the photodissociation
reaction $^{148}$Gd\rga $^{144}$Sm and the inverse capture reaction
$^{144}$Sm\rag $^{148}$Gd. At the end of this section, some remarks on
odd nuclei are given.

The most effective energy $E_{\rm{eff}}$ for \rga\ reactions has
already been defined in Eq.~(\ref{eq:gamow_alpha}). Compared to the
\rag\ capture reaction, this energy is shifted by the binding energy
$Q_\alpha$. For simplicity, the shift of $E_{\rm{eff}}$ to lower
energies because of the energy dependence of the S-factor is neglected
in the following section (see Fig.~\ref{fig:gamow_svar} and
Sect.~\ref{sec:present}). 

\begin{figure}[hbt]
\includegraphics[ bb = 20 112 555 422, width = 8.0cm, clip]{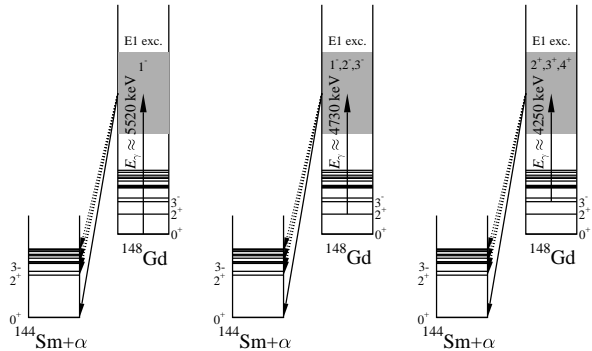}
\caption{
\label{fig:gamow_ga}
Gamow window (shaded area)
for the $^{148}$Gd\rga $^{144}$Sm reaction for the ground
state of $^{148}$Gd ($0^+; 0\,{\rm{keV}}$; left) and for the first excited
states ($2^+; 784\,{\rm{keV}}$; middle; and $3^-; 1273\,{\rm{keV}}$;
right) at a temperature of $T_9 = 2.5$ \cite{MohrTours}. Only the
dominating E1 excitation is shown. See Sect.~\ref{sec:inverse} for
details.
}
\end{figure}

\begin{figure}[hbt]
\includegraphics[ bb = 20 112 555 422, width = 8.0cm, clip]{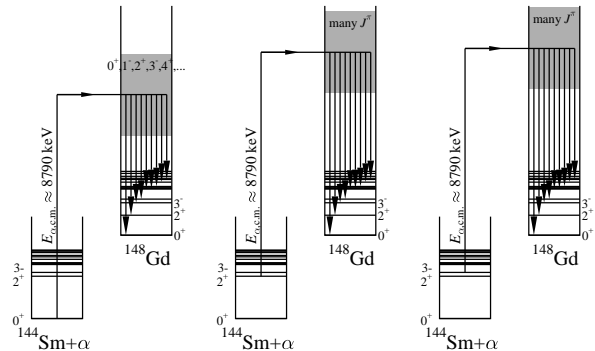}
\caption{
\label{fig:gamow_ag}
Gamow window (shaded area) 
for the $^{144}$Sm\rag $^{148}$Gd reaction for the ground
state of $^{144}$Sm ($0^+; 0\,{\rm{keV}}$; left) and for the first excited
states ($2^+; 1660\,{\rm{keV}}$; middle; and $3^-; 1810\,{\rm{keV}}$;
right) at a temperature of $T_9 = 2.5$. See
Sect.~\ref{sec:inverse} for details. 
}
\end{figure}

In laboratory experiments the target nucleus is in its ground
state. For \rga\ reactions a photon with the energy around
$E_{\rm{eff}}^{(\gamma,\alpha)}$ is required, and the dominating E1
transition leads to the excitation of a $1^-$ state in $^{148}$Gd
which decays dominantly
to the ground state of $^{144}$Sm by $p$-wave $\alpha$ emission 
(see Fig.~\ref{fig:gamow_ga}, left). In general,
$\alpha$ emission at energies below the Coulomb barrier favors the
ground state decay because of the largest tunneling probability for
the highest energy. For
simplicity, the energy and width of the Gamow window are calculated from
Eqs.~(\ref{eq:gamow_alpha}) and (\ref{eq:width}). The slight shift to
lower energies as discussed in Sect.~\ref{sec:present} is neglected
in Figs.~\ref{fig:gamow_ga} and \ref{fig:gamow_ag}.

In a hot stellar environment the nucleus $^{148}$Gd may be thermally
excited. A state at $E_x$ with $J^\pi$ is thermally populated
according to the Boltzmann statistics:
\begin{equation}
\frac{n(E_x,J^\pi)}{n(0,0^+)} = (2J+1) \, \exp{(-E_x/kT)}
\label{eq:thermal}
\end{equation}
The situation is illustrated for the first two excited states of
$^{148}$Gd in Fig.~\ref{fig:gamow_ga} (middle and right). The thermal
excitation reduces the required photon energy
$E_{\rm{eff}}^{(\gamma,\alpha)}$ by $E_x$ leading to the same window
of excitation energies in $^{148}$Gd (Fig.~\ref{fig:gamow_ga}, grey
shaded). Now the dominant E1 transition from the $2^+$ ($3^-$) state
leads to $J^\pi = 1^-, 2^-, 3^-$ ($2^+, 3^+, 4^+$). Most of these
states decay by $\alpha$ emission mainly to the ground state of
$^{144}$Sm (full line) with the exception of the unnatural parity
states with $J^\pi = 2^-, 3^+$ 
decaying to excited states in
$^{144}$Sm (dashed lines).

The contributions of these thermally excited states on the stellar
cross section $\sigma^\ast$ can roughly be estimated. The smaller
occupation probability which scales with $\exp{(-E_x/kT)}$, see
Eq.~(\ref{eq:thermal}), is compensated by the higher photon density at
the relevant effective energy $E_{\rm{eff}}^{(\gamma,\alpha)} - E_x$
which is required for the photodissociation of the excited state at
$E_x$. Consequently, all thermally populated states have comparable
contributions to the stellar cross section
$\sigma^\ast_{(\gamma,\alpha)}$ in Eq.~(\ref{eq:rate}). The ground
state cross section $\sigma^{\rm{lab}}$ and rate $\lambda^{\rm{lab}}$
are usually only small contributions to the stellar cross section
$\sigma^\ast$ and rate $\lambda^\ast$ (see also \cite{Mohr05,Mohr06}).

A precise estimate of the contributions of thermally excited states
requires the E1 $\gamma$-ray strength function at the required energy
$E_{\rm{eff}}^{(\gamma,\alpha)} - E_x$, i.e.~at relatively low
energies far below the giant dipole resonance (GDR). Based on the
Brink-Axel hypothesis, the photon excitation cross section of an
excited state is similar to the photon excitation of the ground state,
but shifted by the energy $E_x$ \cite{Bri55,Axe62,Bar73}. For increasing
$E_x$ the E1 $\gamma$-ray strength decreases at the requested energy
$E_{\rm{eff}}^{(\gamma,\alpha)} - E_x$ because of the increasing
distance to the dominating peak at the GDR. The properties of the
low-energy tail of the GDR are discussed in \cite{Kop90}, and its
astrophysical relevance is analyzed e.g.\ in \cite{Gor98}.

Typically one finds stellar enhancement factors of the order of $100 -
10000$ for photon-induced reactions on heavy nuclei. The stellar
enhancement factor depends on the low-energy behavior of the E1
$\gamma$-ray strength function and the level densities which enter
into statistical model calculations. Stellar enhancement factors for
heavy nuclei are listed in \cite{Uts06,Uts03,Vog01,Uts06a}. Because of
the high level density the influence of electromagnetic selection
rules is relatively small for heavy nuclei compared to their
dominating role in the photodissociation of light nuclei as
discussed in \cite{Mohr06}.

On the other hand, the situation is different for the determination of 
$E_{\rm{eff}}^{(\alpha,\gamma)}$ of the \rag\ capture
reaction. Again, in the laboratory the target nucleus ($^{144}$Sm) is
in its ground state. The effective energy
$E_{\rm{eff}}^{(\alpha,\gamma)}$ is defined by the tunneling through
the Coulomb barrier, e.g., for $^{144}$Sm\rag $^{148}$Gd one obtains
$E_{\rm{eff}}^{(\alpha,\gamma)} = 8.79$\,MeV at $T_9 =2.5$. Levels
with natural parity in $^{148}$Gd around $E_x =
E_{\rm{eff}}^{(\alpha,\gamma)}$ may be populated and subsequently 
decay mainly by
E1 transitions to all low-lying levels of $^{148}$Gd. This is
illustrated in Fig.~\ref{fig:gamow_ag}, left part.

If $^{144}$Sm is thermally excited, the situation is illustrated in
Fig.~\ref{fig:gamow_ag}, middle for $E_x = 1660$\,keV, $J^\pi = 2^+$,
and right for $E_x = 1810$\,keV, $J^\pi = 3^-$. The required energy
for the tunneling through the Coulomb barrier does not change for the
thermally excited $^{144}$Sm; thus, one finds the same effective energy
$E_{\rm{eff}}^{(\alpha,\gamma)} = 8.79$\,MeV as for the ground
state. Levels with many parities may be populated now, but shifted by
$E_x$ to higher energies. These excited states in $^{148}$Gd decay
again to all low-lying states as indicated in Fig.~\ref{fig:gamow_ag}.

Again, the contribution of thermally excited states to the stellar
cross section $\sigma^\ast_{(\alpha,\gamma)}$ can be estimated. The
occupation probability again scales with $\exp{(-E_x/kT)}$, see
Eq.~(\ref{eq:thermal}). But now there is no compensation because the
same energy $E_{\alpha}$ is required for the tunneling,
independent of the excitation energy of $^{144}$Sm. Hence, the
contribution of excited states to the stellar cross section
$\sigma^\ast$ and reaction rate $\lambda^\ast$ is much smaller for
\rag\ reactions compared to \rga\ reactions.

Summarizing the above, a laboratory measurement of the \rag\ capture
cross section $\sigma^{\rm{lab}}_{(\alpha,\gamma)}$ is much closer to
the stellar cross section $\sigma^{\ast}_{(\alpha,\gamma)}$, compared
with a laboratory measurement of the \rga\ photodissociation cross
section $\sigma^{\rm{lab}}_{(\gamma,\alpha)}$ which is a tiny fraction 
of the stellar cross section $\sigma^{\ast}_{(\gamma,\alpha)}$. 
The stellar reaction rates
$\lambda^\ast_{(\alpha,\gamma)}$ of the \rag\ reaction and
$\lambda^\ast_{(\gamma,\alpha)}$ of the \rga\ reaction are linked
together by the detailed balance theorem (e.g.~\cite{Rau00}).    
Therefore, putting aside experimental difficulties for \rag\ 
reactions around the effective energy, \rag\ experiments seem to be 
better suited for the determination of astrophysical 
\rga\ reaction rates than \rga\ experiments.  
However, a careful
consideration of Figs.~\ref{fig:gamow_ga} and \ref{fig:gamow_ag} shows
that the experimental \rga\ cross section consists mainly of a
well-defined transition from the ground state of the target nucleus
via E1 excitation and $\alpha$ emission to the ground state of the
daughter nucleus. This well-defined transition gives the chance for a
precise test of theoretical predictions which cannot be performed in
the same way for \rag\ reactions (except that all $\gamma$-rays of the
\rag\ reaction could be resolved in the \rag\ experiment - which is
quite difficult for heavy nuclei).

Therefore, 
the \rag\ experiment, in principle, provides the best determination of
an individual \astph\ \rag\ and \rga\ reaction rate, but \rga\
experiments provide the most stringent test for the statistical model
and are thus absolutely necessary because of the huge number of reaction
rates to be determined. So both types of experiments are complementary
and should be performed to learn more about stellar \rag\ and \rga\
reaction rates.

Many of the above arguments remain valid for the case of odd
nuclei. However, in nuclei with odd proton or neutron number excited
states are located at lower excitation energies compared to even-even
nuclei. A significant population of these levels is found at typical
\ppro\ temperatures around $T_9 \approx 2 - 3$. These low-lying
excited states have various quantum numbers $J^\pi$. The Gamow window
for \rga\ and \rag\ reactions is the same as for even-even nuclei (see
Figs.~\ref{fig:gamow_ga} and \ref{fig:gamow_ag}); however, it may be
slightly modified by the selection rules for the electromagnetic
excitation process and the additional centrifugal barrier for the
$\alpha$ particle emission.

Contrary to even-even nuclei
-- because of the low-lying excited states --
the stellar \rag\ cross section $\sigma^\ast_{(\alpha,\gamma)}$ and
capture rate $\lambda^\ast_{(\alpha,\gamma)}$ may now be significantly
different from the laboratory measurements for
$\sigma^{\rm{lab}}_{(\alpha,\gamma)}$ and
$\lambda^{\rm{lab}}_{(\alpha,\gamma)}$. Similar to even-even nuclei,
the stellar photodissociation rate $\lambda^\ast_{(\gamma,\alpha)}$
is much larger than the laboratory rate
$\lambda^{\rm{lab}}_{(\gamma,\alpha)}$. The influence of the
electromagnetic selection rules and the additional centrifugal barrier
is difficult to predict in general.

Measurements of laboratory cross sections for \rag\ capture and \rga\
photodissociation reactions are necessary to get the most stringent
restrictions for the theoretical predictions of cross sections and
reaction rates of odd nuclei. Neither a single \rag\
experiment nor a \rga\ experiment is able to provide stellar reaction
rates for odd nuclei. In this case experimental data have to be
completed by theoretical considerations to derive stellar reaction
rates from experimental data.

\subsection{\rpg\ and \rgp\ Reactions}
\label{subsec:rgp2}
The above arguments for odd nuclei are also valid for the relation
between \rgp\ and \rpg\ reactions which are also governed by the
tunneling probability through the Coulomb barrier. Depending on spins
and parities of target and residual nuclei, the influence of
electromagnetic selection rules and the additional centrifugal barrier
may be enhanced in the laboratory especially when the reaction cross
section is strongly suppressed by a strong mismatch
between the spins of the target ground state and the populated
low-lying states in the residual nucleus. As soon as many levels
are thermally populated, it is very unlikely that electromagnetic
selection rules and the centrifugal barrier suppress all possible
transitions for the stellar \rgp\ or \rpg\ cross section or reaction
rate. Again, laboratory measurements of the \rpg\ and \rgp\ reactions
are required although the measured cross sections in the lab may be
only a small contribution of the stellar cross section. Instead, the
experimental data have to provide stringent restrictions for
theoretical calculations.

\subsection{\rng\ and \rgn\ Reactions}
\label{subsec:rgn2}
Contrary to the previous cases, \rng\ and \rgn\ reaction cross
sections show a much 
weaker energy dependence because there is no
repulsive Coulomb interaction. Consequently, the role of the
centrifugal barrier is enhanced, and the ground state contributions to
the stellar reaction rates of both \rng\ and \rgn\ reactions may be
small. Measurements of \rng\ cross sections in the laboratory and the
derived \rng\ rates may be significantly different from stellar rates
at high temperatures around $T_9 = 2 - 3$. However, at typical \spro\
temperatures ($kT \approx 25$\,keV) the 
population of excited states remains small, and stellar \rng\ reaction
rates can usually be determined from laboratory measurements with good
accuracy. At \spro\ temperatures the stellar \rgn\ reaction rate is
negligible because of ($i$) the negligible number of thermal photons
with energies of the order of the neutron separation energy and ($ii$)
the negligible thermal population of excited states in most
cases. Nevertheless, \rgn\ experiments may be used to provide
restrictions for theoretical predictions of the inverse \rng\ capture
cross sections \cite{Sonn03,Mohr04,Shi05}. This is especially
important for unstable \spro\ branching nuclei with short half-lives
which cannot be studied by direct \rng\ capture experiments.

\section{Photodissociation at electron synchrotron based $\gamma$-ray
  sources} 
\label{sec:spring8}
Measurements of photodissociation cross sections are currently
limited to the neutron channel
\cite{MohrPLB,Sonn04,Uts03,Shi05}. Experimental techniques involved in
these measurements are either photoactivation with continuous
bremsstrahlung \cite{MohrPLB,Sonn04} or direct neutron counting with
quasi-monochromatic $\gamma$ rays from laser Compton backscattering
\cite{Uts03,Shi05}. It is a challenge to experimentalists to measure
\rga\ or \rgp\ cross sections of astrophysical significance primarily
because of a lack of appropriate $\gamma$ sources that enables one
to measure the Coulomb-suppressed cross sections. Although a pioneering 
attempt has been made of investigating the $\alpha$ channel for $^{92}$Mo 
\cite{Wag05}, bremsstrahlung may not ideally be suited to determining 
\rga\ or \rgp\ cross sections because unfolding integrated yield curves is 
not straightforward.  Instead of relying on the unfolding procedure, 
a method of superposing bremsstrahlung spectra with different end-point
energies was used to deduce the laboratory photoreaction rate in
Eq.~(\ref{eq:ratelab}) at temperatures of billions of Kelvin
and/or the excitation function of cross sections of $s$-wave nature
\cite{MohrPLB}. 
However, this method suffers from the fact that neither experiment nor 
simulation can determine the shape of the end-point portion of bremsstrahlung 
with high precision \cite{MohrPLB,Sonn04,Vog01}.

\subsection{Blackbody synchrotron radiation at SPring-8}
\label{subsec:sp8}
Recently, it was found \cite{Uts05_NIMA} that the high energy part of
the intense synchrotron radiation produced by a ten-Tesla
superconducting wiggler at SPring-8 well coincides with the stellar
blackbody radiation at temperatures in the range $T_9 = 1.9 - 4.4$
depending on the magnetic field strength from 4 to 10\,T.  This
temperature range of the synchrotron radiation overlaps with that expected
in the \ppro\ nucleosynthesis.
The synchrotron radiation at an
equivalent temperature $T_9 = 4.4$ has the highest $\gamma$ flux that
is experimentally most favorable.  The slight difference of the
experimentally most favorable synchrotron radiation from the upper
limit of the \ppro\ temperature does not hamper the value of
experimental data that constrain the laboratory cross section as
discussed below. 
It is noted that the integrand
$n_{\gamma}(E,T) \, \sigma^{\rm{lab}}_{(\gamma,\alpha)}(E)$ in
Eq.~(\ref{eq:ratelab}) after 
replacing the blackbody radiation $n_{\gamma}(E,T)$ with the
synchrotron radiation can be directly obtained from the experimental
yield (see details below). One of the most important applications is
to determine the  photodissociation rate of $^{180}$Ta$^{\rm{m}}$\rgn
$^{179}$Ta which is of direct relevance to the \ppro\ nucleosynthesis of
$^{180}$Ta$^{\rm{m}}$. It was also shown in \cite{Uts05_NIMA} that the
``blackbody synchrotron radiation'' can be used to study the
$^{181}$Ta\rga $^{177}$Lu reaction by photoactivation. 
In this section, we
perform a comprehensive study of the experimental feasibility of
determining the laboratory \rga\ or \rgp\ reaction rates. 

We remark again that the laboratory reaction rate is a small fraction of the
stellar photoreaction rate, where photoreactions on nuclei in excited
states thermally populated under stellar conditions dominate over the
laboratory rate by a factor of $100 - 10000$ (see
Sect.~\ref{sec:inverse}).
The laboratory rate
can cast new light into the \rga\ cross section
$\sigma^{\rm{lab}}_{(\gamma,\alpha)}(E)$ in Eq.~(\ref{eq:ratelab}),
especially the $\gamma$ transmission coefficient involved in the
Hauser-Feshbach model cross section.  Photonuclear reactions best
probe the E1 $\gamma$-ray strength function which constitutes the main
term of the $\gamma$ transmission coefficient.  Although the
laboratory cross section has a threshold at the particle separation
energy, the laboratory rate can be used to constrain the model E1
$\gamma$ strength function not only above but also below the reaction
threshold.  The E1 $\gamma$ strength function below the reaction
threshold has direct impact on the stellar reaction rate since E1
giant resonance is built on individual excited states according to the
Brink hypothesis \cite{Bri55}.

There is an important feature that helps an experimental study of the
$\alpha$ channel. As pointed out in Sec.~\ref{subsec:basic}, the
reaction becomes exothermic with a positive $Q$-value in the $\alpha$
channel of photoreactions on heavy nuclei: $Q(\gamma,\alpha) > 0$.
Therefore, the most
effective $\gamma$-ray energy in Eq.~(\ref{eq:gamow_alpha}) for heavy
nuclei decreases by the amount of $Q_\alpha = - Q(\gamma,\alpha)$.  As a
result, the required $\gamma$ energy is comparable to or even smaller
than that for the neutron channel (see Fig.~\ref{fig:gamow_svar}), while
the energy of $\alpha$ particles is boosted by the exothermic
reaction. The $\gamma$ flux of the synchrotron radiation is very high
in the low energy region, which may overcome small \rga\
cross sections governed by the Coulomb barrier.  On the other hand,
most of \rga\ reactions on stable nuclei result in the
production of stable residual nuclei so that an experimental technique
of direct particle counting needs to be applied. In this respect, the
exothermic reaction helps to detect the promptly 
emitted $\alpha$ particles.

\subsection{Expected \rga\ yields at SPring-8}
\label{subsec:sp8-yield}
We discuss below measurements of the laboratory \rga\ rate with the
synchrotron radiation produced by the 10-T superconducting wiggler
at SPring-8. It is foreseen to propose a dedicated beamline for
experiments with the blackbody synchrotron radiation (BSR).

The number of $\alpha$ particles $N$ produced per second by irradiating
a target sample with BSR is expressed by
\begin{equation}
N(t) = n_{T} \int\limits_{S_{\alpha}}^{\infty} n_{\gamma}^{BSR}(E,T) \,
\sigma^{\rm{lab}}_{(\gamma,\alpha)}(E) \, dE,
\label{eq:EventRate}
\end{equation}
where $n_T$ is the number of target nuclei per unit area included in
the target sample and $n_{\gamma}^{BSR}(E,T)$ is the flux of BSR 
[s$^{-1}$ MeV$^{-1}$] with an equivalent
blackbody temperature $T$.  One can see that it is straightforward to
convert this experimental quantity to the laboratory reaction rate
in Eq.~(\ref{eq:ratelab}).

\begin{figure}[htb]
\includegraphics[ bb = 50 200 510 580, width = 8.0cm, clip]{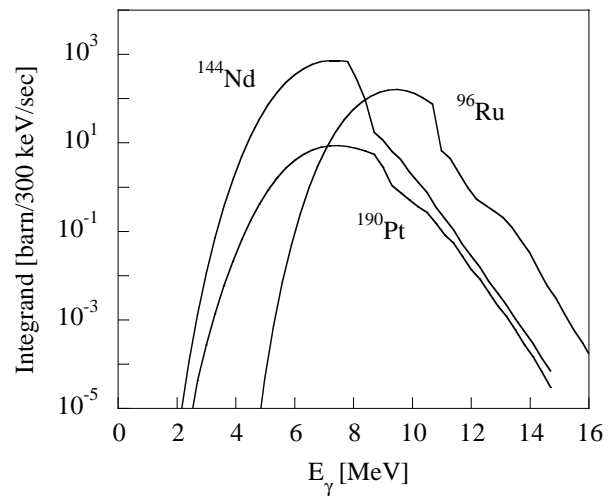}
\caption{
\label{fig:PtNdRu}
Integrand of Eq.~(\ref{eq:EventRate}) for \rga\ reactions on
$^{190}$Pt, $^{144}$Nd, and $^{96}$Ru at $T_9 = 4.4$ corresponding to
the maximum magnetic field of 10\,T.
}
\end{figure}

By using the $\gamma$-ray flux (Fig.~3 of Ref.~\cite{Uts05_NIMA}) and
the \rga\ cross section of \cite{Rau04}, the event rate of
Eq.~(\ref{eq:EventRate}) was calculated for all the 233 reactions
compiled in \cite{Rau04}. Fig.~\ref{fig:PtNdRu} shows the integrand of
Eq.~(\ref{eq:EventRate}) for \rga\ reactions on $^{190}$Pt,
$^{144}$Nd, and $^{96}$Ru for $T_9 = 4.4$ corresponding to the maximum
magnetic field of 10\,T. (The same examples have been chosen as in the
previous Sects.~\ref{sec:present} and \ref{sec:inverse} except that the
unstable $^{148}$Gd has been replaced 
by the experimentally most favorable case $^{144}$Nd.) 
The integrand shows a peak at 7.50\,MeV for
$^{190}$Pt, at 7.20\,MeV for $^{144}$Nd, and at 9.49\,MeV for $^{96}$Ru.
The most probable $\alpha$ energy is 10.75\,MeV for $^{190}$Pt ($Q_\alpha =
-3.25$\,MeV), at 9.10\,MeV for $^{144}$Nd ($Q_\alpha = -1.90$\,MeV),
and at 7.80\,MeV 
for $^{96}$Ru ($Q_\alpha = 1.69$\,MeV). Note that $Q_\alpha = -
Q(\gamma,\alpha)$.

The integrand in Eq.~(\ref{eq:EventRate}) and Fig.~\ref{fig:PtNdRu} exhibits a
sudden decrease at an energy corresponding to the neutron separation
energy, i.e., at 8.9\,MeV for $^{190}$Pt, 7.8\,MeV for $^{144}$Nd, and
at 10.7\,MeV for $^{96}$Ru; as soon as the neutron channel opens,
$\alpha$ emission competes with neutron emission which is not
suppressed by the Coulomb barrier. A similar behavior is also found
for the \rag\ cross sections, see Fig.~\ref{fig:sfaktvar}.
 
The results assure an experimental
feasibility when both the number of target nuclei for the event rate
and the kinetic energy for the $\alpha$ detection are taken into
account as follows.  When a target foil of a thickness corresponding
to half the range of $\alpha$ particles with the most 
probable energy
is used, the event rate is 44\,cph (counts/hour) for $^{190}$Pt,
2200\,cph for $^{144}$Nd, and 540\,cph for $^{96}$Ru.  These event rates are
sufficient from the viewpoint of the direct detection of $\alpha$
particles. Several foils can be mounted inside a vacuum chamber for a
simultaneous irradiation with BSR and $\alpha$
particles emitted can be detected with suitable detectors, for
example, arrays of silicon detectors each surrounding the individual
target foil. Table \ref{tab:exp} lists experimental conditions for
\rga\ reactions with the event rate at the level of 10\,cph
or more. The count rate estimates are based on 100\,\% enriched
targets. Highly enriched targets are necessary for the suggested
experiments because the detected
$\alpha$ particle energy does not allow to distinguish among
\rga\ reactions on different nuclei in the target.
One sees that the synchrotron radiation from the 10-T
superconducting wiggler at SPring-8 can provide an unprecedented
experimental opportunity to measure \rga\ cross sections
on many nuclei in the \ppro\ path.  It is to be noted that besides the 
event rate, the experimental feasibility, as a matter of course, 
depends also on the availability, the enrichment, and the cost of
target foils. Finally, the experimental background conditions have to
be analyzed carefully because of the high intensity of 
low-energy photons in the BSR spectrum.

Similar calculations for \rgp\ reactions show that there is an experimental
opportunity for 11 \rgp\ reactions on nuclei from $^{59}$Co to
$^{89}$Y at SPring-8 (not listed here).

\begingroup
\squeezetable
\begin{table*}
\caption{ 
\label{tab:exp}
Conditions for \rga\ reactions that can be experimentally studied with
the synchrotron radiation from the 10-T superconducting wiggler at
SPring-8.  The event rate is given for a single target-foil of the
thickness corresponding to a half the range of $\alpha$ particles in
the target foil. The table is separated into three subtables with
rates below 100\,cph, above 100\,cph, and above 1000\,cph.  
}
\begin{tabular}{cccccc}
\hline\hline
Target nucleus & Reaction Q-value & Gamow peak  & Integral & Target thickness & Event rate \\ \hline
 & [MeV] & [MeV] & [barn/sec] & [$\mu$m] & counts per hour \\ \hline 
$^{69}$Ga & -4.5 & 9.9 & 1.0E+2  & 18 & 35  \\
$^{72}$Ge &  -4.5 & 10.2 & 4.9E+1 & 22 & 17  \\
$^{93}$Nb & -1.9 & 8.1 & 2.5E+1 & 8.0 & 8  \\
$^{96}$Mo & -2.8 & 9.1 & 2.0E+1 & 8.5 & 9  \\
$^{98}$Ru & -2.2 & 10.0 & 1.5E+2 & 7.8 & 62  \\  
$^{102}$Pd & -2.2 & 10.2 & 9.7E+1 & 9.0 & 44  \\
$^{106}$Cd & -1.6 & 10.0 & 1.8E+2 & 13 & 87  \\
$^{112}$Sn & -1.8 & 10.2 & 3.0E+1 & 16 & 14  \\
$^{122}$Te & -1.1 & 9.5 & 2.2E+1 & 19 & 9  \\
$^{130}$Ba & -0.52 & 9.5 & 3.6E+1 & 38 & 16  \\
$^{145}$Nd & +1.6 & 6.3 & 9.3E+1 & 16 & 32  \\
$^{149}$Sm & -1.2 & 9.4 & 1.1E+2 & 16 & 40  \\
$^{151}$Eu & +2.0 & 6.0 & 1.4E+2 & 15 & 32  \\
$^{154}$Gd & +0.92 & 8.7 & 1.57E+1 & 21 & 7  \\
$^{156}$Dy & +1.8 & 7.8 & 1.4E+2 & 20 & 66  \\
$^{162}$Er & +1.6 & 8.1 & 4.8E+1 & 20 & 23  \\
$^{168}$Yb & +2.0 & 8.6 & 5.0E+1 & 30 & 27  \\
$^{170}$Yb & +1.7 & 7.4 & 1.9E+1 & 24 & 8  \\
$^{174}$Hf & +2.5 & 7.5 & 1.3E+2 & 15 & 65  \\
$^{176}$Hf & +2.3 & 8.3 & 4.1E+1 & 16 & 22  \\
$^{178}$Hf & +2.1 & 7.5 & 1.6E+1 & 14 & 7  \\
$^{180}$Ta & +2.0 & 6.9 & 2.3E+1 & 10 & 9  \\
$^{180}$W & +2.5 & 7.8 & 3.8E+1 & 11 & 20  \\
$^{184}$Os & +3.0 & 7.5 & 8.7E+1 & 10 & 46  \\
$^{186}$Os & +2.8 & 7.5 & 4.1E+1 & 10 & 21  \\
$^{190}$Pt & +3.3 & 7.5 & 8.2E+1 & 11 & 44  \\ \hline
$^{66}$Zn & -4.6 & 10.0 & 8.7E+2 & 7 & 283  \\
$^{70}$Ge & -4.1 & 9.8 & 1.3 E+3 & 11 & 459  \\
$^{74}$Se & -4.1 & 10.1 & 3.7E+2 & 13 & 134  \\
$^{94}$Mo & -2.1 & 9.6 & 7.2E+2 & 5 & 170  \\
$^{96}$Ru & -1.7 & 9.5 & 1.2E+3 & 8 & 543  \\
$^{120}$Te & -0.3 & 8.7 & 3.7E+2 & 19 & 156  \\
$^{142}$Ce & +1.3 & 7.2 & 1.5 E+3 & 19 & 566  \\
$^{147}$Sm & +2.3 & 6.3 & 8.7E+2 & 18 & 346  \\
$^{148}$Sm & +2.0 & 7.2 & 1.6E+3 & 20 & 697  \\
$^{150}$Sm & +1.4 & 7.8 & 3.2E+2 & 20 & 140  \\ 
$^{152}$Gd & +2.2 & 7.2 & 1.7E+3 & 20 & 763  \\ \hline
$^{64}$Zn & -4.0 & 9.4 & 3.1E+3 & 7 & 1046  \\
$^{144}$Nd & +1.9 & 7.2 & 5.1E+3 & 20 & 2227  \\ \hline\hline
\end{tabular}
\end{table*}
\endgroup

\section{Conclusion and Recommendations}
\label{sec:conc}
The current status of \astph ly relevant photon-induced cross sections
and reaction rates was discussed in detail with a focus on \rga\
reactions where almost no experimental data are available in the
energy range of the astrophysical \ppro . It is shown that
experimental \rag\ data, in principle, provide the best estimate 
for stellar \rga\ reaction rates though measurements of \rag\ reactions
around the effective energy below the Coulomb barrier themselves are 
very difficult.  Although
experimental \rga\ data in the laboratory are not perfectly suited for a direct
determination of stellar \rga\ reaction rates, such data nevertheless
provide best insight into particular ingredients of the required
statistical model calculations, namely the $\gamma$-ray
strength function, the level density, and the $\alpha$-nucleus potential.

It is found that the energy dependence of the widely used \astph\
S-factor $S(E)$ hampers the extrapolation of cross sections to low
energies especially for the heavy nuclei under study in this work. As
a result, the most effective energy $E_{\rm{eff}}$ is slightly lower
than estimated from the usual formula in
Eq.~(\ref{eq:gamow_alpha}). An improved S-factor $S_C(E)$ is
suggested which is based on the Coulomb wave functions instead of the
simplistic Gamow factor $\exp{(-2 \pi \eta)}$. The energy dependence of
$S_C(E)$ is much smaller than for $S(E)$; thus, extrapolations using
$S_C(E)$ should be more reliable.

The high-energy part of synchrotron radiation produced by a 10-T 
superconducting wiggler at SPring-8 is an excellent tool to
mimic the blackbody radiation of temperatures around $T_9 = 1.5 - 4.4$
which are typical or slightly higher than \ppro\ conditions. 
Based on the statistical model cross sections of \cite{Rau04} and 
the photon intensity of \cite{Uts05_NIMA}, it is shown that 
a noticeable number of \rga\ cross
sections can be measured with reasonable count rates. These new
experimental results are expected to lead to a significant improvement
of nuclear parameters involved in the statistical model, in particular,
$\gamma$-ray strength functions and $\alpha$-nucleus potentials,
and thus may help to reduce the uncertainties of \ppro\
nucleosynthesis calculations.

\begin{acknowledgments}
We thank T.\ Rauscher for providing his calculations of the
$^{148}$Gd\rga $^{144}$Sm reaction and S.\ Goriely for his comments on
the quantitative aspects of the stellar reaction rate. This work was
supported partly by OTKA (grants T042733 and T068801). Zs.~F.\
acknowledges support from the Bolyai grant. H.~U.\ acknowledges support
from the Japan Private School Promotion Foundation and the Japan
Society of the Promotion of Science.
\end{acknowledgments}

\end{document}